\documentclass[a4paper]{article}
\usepackage{amssymb}
\title{The Nature of Individual Choices}
\author{Redjan F. Shabani\\Master Student\\Artificial Intelligence \& Robotics\\University of Rome "La Sapienza"\\shabani.1013173@studenti.uniroma1.it}
\date{\today}

\begin{document}
\maketitle

\abstract
In the theory of social choice the research is focused around the projection of individual preference orders to the social preference order. Also, the justification of the preference order formalism begins with the concept of utility i.e. an alternative is preferred to another one if the utility over the first is higher then the utility over the second. In this paper is proposed an ideal model of measuring utilities by considering individuals and alternatives no more as atomic concepts but as being composed by other entities.

\section{Basic definitions}
Individuals, along their existence, tries to reach a series of objectives that, according to them, ensure a satisfiable individual welfare. When an individual judges over an alternative considers his own objectives and observes the possibility that this particular alternative will allow him to reach these objectives.

\newtheorem{defn1}{Definition}
\begin{defn1}
Given a society $V=\{V_1,V_2,...,V_N\}$ and an environment $A=\{A_1,A_2,...,A_M\}$, we define the \emph{universe of the objectives}, the set:
\begin{displaymath}
\Gamma=\{\gamma_1,\gamma_2,...,\gamma_R\}
\end{displaymath}
as the set of all the required objectives by the individuals in $I$ and the objectives offered by the alternatives in $A$.
\end{defn1}

For each alternative $A_m \in A$ we can put in evidence the set of the offered objectives, that for simplicity we will give it the same name as the alternative itself:
\begin{displaymath}
A_m=\{\alpha_1^m,\alpha_2^m,...,\alpha_H^m\}
\end{displaymath}
where $\alpha_h^m \in \Gamma$ for all $h=1,...,H$.\\
Similarly, each individual $V_n \in V$ is characterised by the set of the required objectives:
\begin{displaymath}
V_n=\{\nu_1^n,\nu_2^n,...,\nu_K^n\}
\end{displaymath}
where $\nu_k^n \in \Gamma$ for all $k=1,...,K$.\\
Considering the universe $\Gamma$ as a-priori entity, we can treat a given alternative or individual as an element of $\wp(\Gamma)$\footnote{$\wp(\Gamma)$ is the set of all subsets of $\Gamma$}
The set of all the alternatives definable over $\Gamma$ is simply the following:
\begin{displaymath}
A^*=\{A_m\mid A_m\in\wp(\Gamma)-\phi\}
\end{displaymath}
Similarly, the set of all individuals definable over $\Gamma$ is:
\begin{displaymath}
V^*=\{V_n\mid V_n\in\wp(\Gamma)-\phi\}
\end{displaymath}

\newtheorem{defn2}[defn1]{Definition}
\begin{defn2}
Given an environment $A$, we define the opportunity universe as the following  set:
\begin{displaymath}
\Gamma_A=\bigcup_{m=1}^M A_m
\end{displaymath}
containing all the objectives offered by at least one alternative.
\end{defn2}

\newtheorem{defn3}[defn1]{Definition}
\begin{defn3}
Given a society $V$, we define the exigence universe as the following  set:
\begin{displaymath}
\Gamma_V=\bigcup_{n=1}^N V_n
\end{displaymath}
containing all the objective required by at least one individual.
\end{defn3}
We can distinguish the following subset of the universe $\Gamma$:
\begin{itemize}
\item $\Gamma_A-\Gamma_V$, not offered request
\item $\Gamma_V-\Gamma_A$, not requested offer
\item $\Gamma_A\cap\Gamma_V$, requested offer and offered request
\end{itemize}

\section{Utility measure based on the set cardinality}

The next step is to define an utility measure. An immediate measure may be the cardinality between the individual set of objectives and the alternative set of objectives. We will use the notation $u(A_m\mid V_n)$ or $u_n(A_m)$ to denote the utlity of $V_n$ over $A_m$.
\newtheorem{defn2.1}[defn1]{Definition}
\begin{defn2.1}
Given $A_m\in A$ and $V_n\in V$ the \emph{cardinal utility} is defined as:
\begin{displaymath}
cu(A_m\mid V_n)=card(A_m\cap V_n)
\end{displaymath}
\end{defn2.1}
This utility measure, if measurable, allows us to order alternatives in decreasing order, but there is a problem; this measure may assume arbitrary values. Suppose we have two individuals such that $V_p=\{\alpha\}$ and $V_q=\{\alpha,\beta,\gamma\}$ an suppose they have to evaluate an alternative characterised by $A_m=\{\alpha,\beta,\gamma\}$. By using the cardinality measure we have $u(A_m\mid V_P)=1$ and $u(A_m\mid V_q)=3$. Both $V_p$ and $V_q$ are satisfied of the choice they have made, because $A_m$ offers all they requested, but the utility are different, this because $V_p$ is less exigent than $V_q$. If we want to homogenise the utility measure we ha define a more specialised cardinal utility.
\newtheorem{defn2.2}[defn1]{Definition}
\begin{defn2.2}
Given $A_m\in A$ and $V_n\in V$, the \emph{normalised cardinal utility} is defined as:
\begin{equation}\label{eq:ncu}
ncu(A_m\mid V_n)=\frac{card(A_m\cap V_n)}{card(V_n)}
\end{equation}
where $ncu(A_m\mid V_n)$ indicates the utility of $V_n$ over $A_m$.
\end{defn2.2}
Let's try to specialise further the utility measure. Individuals, in general, do not treat their own objectives with the same importance. We can suppose that it is assigned a weight for each of the objectives. The individuals have to be characterised by a membership function, like in the fuzzy sets. The classic set definition can be formulated in terms of membership function.\\
GIven a set S, for all $u$ in the universe $U$, the membership function $\mu_S:U\rightarrow\{0,1\}$ of the set $S$ is defined as follows:
\begin{itemize}
\item $\forall u\in S$, $\mu_S(u)=1$
\item $\forall u\notin S$, $\mu_S(u)=0$
\end{itemize}
Fuzzy sets are defined simply by changing the definition of the membership function from $\mu_S:U\rightarrow\{0,1\}$ to $\mu_S:U\rightarrow [0,1]$. An element $u\in U$ can belong to $S$ with a grade of membership less than $1$.\\
Let's return back to our question of defining a cardinal-oriented utility measure .
\newtheorem{ass}{Assumption}
\begin{ass}
Given the universe of the objectives $\Gamma$, for all individual $V_n$ in the society $V$ it is defined a membership function:
\begin{displaymath}
\mu_{V_n}:U \rightarrow [0,1]
\end{displaymath}
\end{ass}
Now is possible to define a further version of the utility measure.
\newtheorem{defn2.3}[defn1]{Definition}
\begin{defn2.3}
Given an individual $V_n\in V$ and an alternative $A_m\in A$ and supposing that the $V_n$ is characterised by the membership function $\mu_n:\Gamma\rightarrow [0,1]$, the fuzzy utility measure $fu(A_m\mid V_n)=fu_n(A_m)$ of $V_n$ over $A_m$ is defined as:
\begin{equation}\label{eq:fcu}
fu(A_m\mid V_n)=\frac{\sum_{\gamma\in A_m} \mu_n(\gamma)}{\sum_{\gamma\in \Gamma}\mu_n(\gamma)}
\end{equation}
\end{defn2.3}

\section{Evaluation Processes}
Once we have a model of how individuals express the own opinions regarding the alternatives, remains to define formally the voting process. Here we will call the entire decision process of the society with the term evaluation process, principally to distinguish from the classical choice process that operates over preference orders.\\
Given an utility measure $u(A_m\mid V_n)=u_n(A_m)$, according (\ref{eq:ncu}), (\ref{eq:fcu}) \footnote{The utility measure can be defined also in another way, independently to the formalism proposed here.}, for each individual $V_n$, we define the \emph{individual profile} as the following vector:
\begin{displaymath}
\mathbf{u}^{(n)}=
\left[
\begin{array}{clcr}
u_n(A_1)\\
u_n(A_2)\\
\vdots \\
u_n(A_M)\\
\end{array}
\right]
\end{displaymath}

\newtheorem{defn3.1}[defn1]{Definition}
\begin{defn3.1}
An \emph{evaluation process} is defined as quadruple:
\begin{displaymath}
E=\langle A,V,U,f \rangle
\end{displaymath}
where:
\begin{itemize}
\item $A=\{A_1,...,A_M\}$ is a finite and non empty set of alternatives, called environment
\item $V=\{V_1,...,V_N\}$ is a finite and non empty set of individuals, called society
\item $U=\{\mathbf{u}^1,...,\mathbf{u}^N\}$ is a set of individual profiles, where $u^n$ is the profile of $V_n$.
\item $f:[0,1]^N\rightarrow[0,1]$ is a function, called evaluation function, that returns a corresponding social profile, given the individual profiles.
\end{itemize}
\end{defn3.1}
The evaluation function is not the same as the social welfare function; the second return the social preference order, given the individual preference orders, while the evaluation function operates over vectors in $\mathbb{R}^M$.

Evaluation processes are characterised by the nature of the evaluation function. An immediate example of evaluation function is the mean of individual profiles. Given an environment $A=\{A_1,...,A_M\}$ and a society $V=\{V_1,...,V_N\}$ the \emph{mean social profile} is defined as the following vector in $\mathbb{R}^M$
\begin{displaymath}
\mathbf{u}=\frac{1}{N}\sum_{n=1}^{N}\mathbf{u}^n
\end{displaymath}.


\begin{thebibliography}{99}
\bibitem {TG}J. V. Neumann, O.Morgensten, \emph{Theory of games and economic behavior}, Princeton University Press, 1944
\bibitem {AR}K. J. Arrow, \emph{Social Choice and Individual Values}, John Wiley \& Sons, 1963
\bibitem {HS}K. J. Arrow, A. K. Sen, \emph{Handbook of social choice and welfare}, Gulf Professional Publications, 2002
\bibitem {FL}L. A. Zadeh, \emph{Fuzzy Sets},Elsevier - Information and Control Vol.8 Issue 3, June 1965, 338-353
\bibitem {RF}R. F Bodley, \emph{Reformulating decision theory using fuzzy set theory and Shafer's theory of evidence}, Elsevier - Fuzzy Sets ans Systems Vol. 139 Issue 2, 2003, 243-266
\end{thebibliography}
\end{document}